\documentclass{revtex4}

\usepackage{graphics}
\usepackage{epsfig}

\begin{document}

\title{New Results from the Muon $g-2$ Experiment}

\author{
E.P.~Sichtermann$^{11}$,
G.W.~Bennett$^{2}$,
B.~Bousquet$^{9}$,
H.N.~Brown$^2$,
G.~Bunce$^2$,
R.M.~Carey$^1$,
P.~Cushman$^{9}$,
G.T.~Danby$^2$,
P.T.~Debevec$^7$,
M.~Deile$^{11}$,
H.~Deng$^{11}$,
W.~Deninger$^7$,
S.K.~Dhawan$^{11}$,
V.P.~Druzhinin$^3$,
L.~Duong$^{9}$,
E.~Efstathiadis$^1$,
F.J.M.~Farley$^{11}$,
G.V.~Fedotovich$^3$,
S.~Giron$^{9}$,
F.E.~Gray$^7$,
D.~Grigoriev$^3$,
M.~Grosse-Perdekamp$^{11}$,
A.~Grossmann$^6$,
M.F.~Hare$^1$,
D.W.~Hertzog$^7$,
X.~Huang$^1$,
V.W.~Hughes$^{11}$,
M.~Iwasaki$^{10}$,
K.~Jungmann$^5$,
D.~Kawall$^{11}$,
B.I.~Khazin$^3$,
J.~Kindem$^{9}$,
F.~Krienen$^1$,
I.~Kronkvist$^{9}$,
A.~Lam$^1$,
R.~Larsen$^2$,
Y.Y.~Lee$^2$,
I.~Logashenko$^{1,3}$,
R.~McNabb$^{9}$,
W.~Meng$^2$,
J.~Mi$^2$,
J.P.~Miller$^1$,
W.M.~Morse$^2$,
D.~Nikas$^2$,
C.J.G.~Onderwater$^7$,
Y.~Orlov$^4$,
C.S.~\"{O}zben$^2$,
J.M.~Paley$^1$,
Q.~Peng$^1$,
C.C.~Polly$^7$,
J.~Pretz$^{11}$,
R.~Prigl$^{2}$,
G.~zu~Putlitz$^6$,
T.~Qian$^{9}$,
S.I.~Redin$^{3,11}$,
O.~Rind$^1$,
B.L.~Roberts$^1$,
N.~Ryskulov$^3$,
P.~Shagin$^9$,
Y.K.~Semertzidis$^2$,
Yu.M.~Shatunov$^3$,
E.~Solodov$^3$,
M.~Sossong$^7$,
A.~Steinmetz$^{11}$,
L.R.~Sulak$^{1}$,
A.~Trofimov$^1$,
D.~Urner$^7$,
P.~von~Walter$^6$,
D.~Warburton$^2$,
and
A.~Yamamoto$^8$.
\\
(Muon $g-2$ Collaboration)\vspace{1ex}
}
\affiliation{
\mbox{$\,^1$Department of Physics, Boston University, Boston, Massachusetts 02215}\\
\mbox{$\,^2$Brookhaven National Laboratory, Upton, New York 11973}\\
\mbox{$\,^3$Budker Institute of Nuclear Physics, Novosibirsk, Russia}\\
\mbox{$\,^4$Newman Laboratory, Cornell University, Ithaca, New York 14853}\\
\mbox{$\,^5$ Kernfysisch Versneller Instituut, Rijksuniversiteit Groningen, NL 9747\,AA Groningen, The Netherlands}\\
\mbox{$\,^6$ Physikalisches Institut der Universit\"at Heidelberg, 69120 Heidelberg, Germany}\\
\mbox{$\,^7$ Department of Physics, University of Illinois at Urbana-Champaign, Illinois 61801}\\
\mbox{$\,^8$ KEK, High Energy Accelerator Research Organization, Tsukuba, Ibaraki 305-0801, Japan}\\
\mbox{$\,^{9}$Department of Physics, University of Minnesota,Minneapolis, Minnesota 55455}\\
\mbox{$\,^{10}$ Tokyo Institute of Technology, Tokyo, Japan}\\
\mbox{$\,^{11}$ Department of Physics, Yale University, New Haven, Connecticut 06520}
}

\begin{abstract}
The Muon $g-2$ collaboration has measured the anomalous magnetic $g$ value,
$a = (g-2)/2$, of the positive muon with an unprecedented uncertainty of
0.7\,parts per million.
The result
\mbox{$a_{\mu^+}(\mathrm{expt}) = 11\,659\,204(7)(5)\ \times 10^{-10}$},
based on data collected in the year 2000 at Brookhaven National Laboratory,
is in good agreement with the preceding data on $a_{\mu^+}$ and $a_{\mu^-}$.
The measurement tests standard model theory, which at the level of the current
experimental uncertainty involves quantum electrodynamics, quantum
chromodynamics, and electroweak interaction in a significant way.
\end{abstract}

\maketitle

\section{Introduction}
The magnetic moment $\vec{\mu}$ of a particle with charge $e$, mass $m$,
and spin $\vec{s}$ is given by
\begin{equation}
  \vec{\mu} = \frac{e}{2mc}\, g\,\vec{s}
\end{equation}
in which $g$ is the gyromagnetic ratio.
For point particles with spin-$\frac{1}{2}$, Dirac theory predicts $g=2$.

Precision measurements of the $g$ values of leptons and baryons have
historically played an important role in the development of particle theory. 
The $g$ value of the proton, for example, is found to differ sizeably from
$2$, which provides evidence for a rich internal proton (spin) structure. 
The lepton $g$ values deviate from $2$ only by about one part in a thousand,
consistent with current evidence that leptons are point particles with
spin-$\frac{1}{2}$.
The anomalous magnetic $g$ value of the electron, \mbox{$a_e = (g_e - 2)/2$},
is among the most accurately measured quantities in physics, and is
presently known with an uncertainty of about four parts per billion
(ppb)~\cite{VanDyck:1987ay}.
The value of $a_e$ is described in terms of standard model (SM) field
interactions, in which it has a leading order contribution of $\alpha/(2\pi)$ from
the so-called Schwinger term.
Nearly all of the measured value is contributed by QED processes involving
virtual electrons, positrons, and photons.
Particles more massive than the electron contribute only at the level of the present experimental uncertainty.

The anomalous magnetic $g$ value of the muon, $a_\mu$, is more sensitive than
$a_e$ to processes involving particles more massive than the electron,
typically by a factor $(m_\mu/m_e)^2 \sim 4 \cdot 10^4.$
A series of three experiments at CERN~\cite{cern-1,cern-2,cern-3} measured
$a_\mu$ to a final uncertainty of about 7~parts per million (ppm), which is
predominantly of statistical origin.
The CERN generation of experiments thus tested electron-muon universality
and established the existence of a $\sim 59$\,ppm hadronic contribution
to $a_\mu$.
Electroweak processes are expected to contribute to $a_\mu$ at the level
of 1.3\,ppm, as are many speculative extensions of the SM.

The present muon $g-2$ experiment at Brookhaven National Laboratory (BNL)
has determined $a_{\mu^+}$ of the positive muon with an uncertainty of
0.7\,ppm from data collected in the year 2000, and aims for a similar
uncertainty on $a_{\mu^-}$ of the negative muon from measurements in
2001.
The continuation of the experiment by a single running period, if funded,
 should bring the design goal precision of 0.4\,ppm within reach.

\section{Experiment}
The concept of the experiment at BNL is the same as that of the last of
the CERN experiments~\cite{cern-1,cern-2,cern-3} and involves the study
of the orbital and spin motions of polarized muons in a magnetic storage
ring.
\begin{figure}
  \begin{center}
    \includegraphics[width=0.60\textwidth]{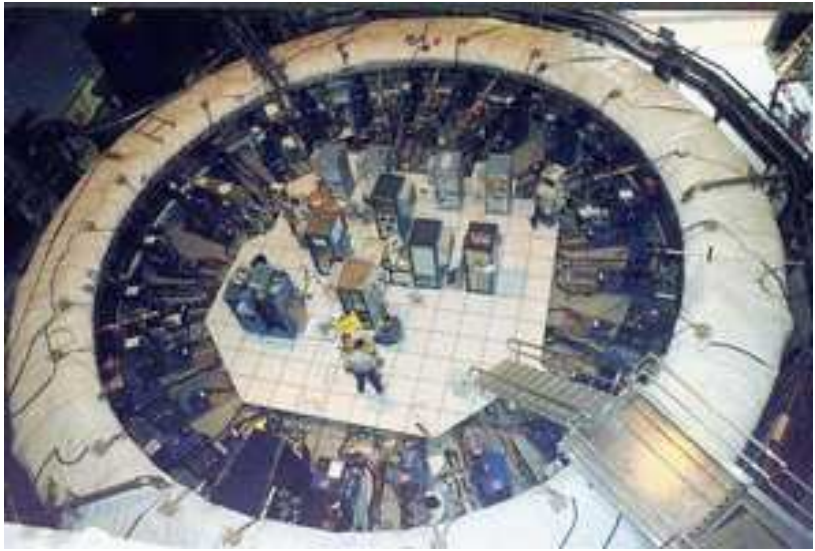}
  \end{center}
  \caption{Top view of the $g-2$ apparatus.  The beam of longitudinally polarized muons enters the superferric storage ring magnet through a superconducting inflector magnet located at 9 o'clock and circulates clockwise after being placed onto stored orbit with three pulsed kickers modules in the 12 o'clock region.  Twenty-four lead scintillating-fiber calorimeters on the inner, open side of the C-shaped ring magnet are used to measure muon decay positrons.
The central platform supports the power supplies for the four electrostatic quadrupoles and the kicker modules.}
  \label{fig:apparatus}
\end{figure}

The present experiment (Fig.~\ref{fig:apparatus}) is situated at the Alternating
Gradient Synchrotron (AGS), which in the year 2000 delivered up to $60 \times 10^{12}$
protons in twelve 50\,ns (FWHM) bunches over its 3\,s cycle.
The 24\,GeV protons from the AGS were directed onto a rotating,
water-cooled nickel target.
Pions with energies of 3.1\,GeV emitted from the target were captured
into a 72\,m straight section of focusing-defocusing magnetic quadrupoles,
which transported the parent beam and naturally polarized muons from
forward pion decays.
At the end of the straight section, the beam was momentum-selected and
injected into the 14.2\,m diameter storage ring magnet~\cite{magnet}
through a field-free inflector~\cite{inflector} region in the magnet
yoke.
A pulsed magnetic kicker~\cite{kicker} located at approximately one quarter
turn from the inflector region produced a 10\,mrad deflection which placed
the muons onto stored orbits.
Pulsed electrostatic quadrupoles~\cite{quads} provided vertical focusing.
The magnetic dipole field of about 1.45\,T was measured with an NMR
system~\cite{nmr} relative to the free proton NMR frequency $\omega_p$
over most of the 9\,cm diameter circular storage aperture.
Twenty-four electromagnetic calorimeters~\cite{calo} read out by
400\,MHz custom waveform digitizers (WFD) were used on the open, inner
side of the C-shaped ring magnet to measure muon decay positrons.
The decay violates parity, which leads to a relation between the
muon spin direction and the positron energy spectrum in the laboratory
frame.
For positrons above an energy threshold $E$, the muon-decay time-spectrum
\begin{equation}
  N(t) = N_0(E) \exp\left(\frac{-t}{\gamma\tau}\right)
       \left[ 1 + A(E) \sin \left( \omega_a t + \phi(E) \right) \right],
  \label{eq:precession}
\end{equation}
in which $N_0$ is a normalization, $\gamma\tau \sim 64\,\mu\mathrm{s}$
is the dilated muon lifetime, $A$ an asymmetry factor, $\phi$ a phase,
and $\omega_a$ the angular difference frequency of muon spin precession
and momentum rotation.
In our measurements, the NMR and WFD clocks were phase-locked to the
same LORAN-C~\cite{loran} frequency signal.

The muon anomalous magnetic $g$ value is evaluated from the ratio of the
measured frequencies, $R = \omega_a/\omega_p$, according to:
\begin{equation}
  a_\mu = \frac{R}{\lambda - R},
\end{equation}
in which $\lambda = \mu_\mu / \mu_p$ is the ratio the muon and proton
magnetic moments.
The value with smallest stated uncertainty,
$\lambda = \mu_\mu / \mu_p =  3.183\,345\,39(10)$~\cite{pdg},
results from measurements of the microwave spectrum of ground state
muonium~\cite{liu} and theory~\cite{kino,nio}.

Important improvements made since our preceding
measurement~\cite{Brown:2001mg} include:
the operation of the AGS with 12 beam bunches,
which contributed to a 4-fold increase in the data collected;
a new superconducting inflector magnet, which greatly improved
the homogeneity of the magnetic field in the muon storage region; 
a sweeper magnet in the beamline, which reduced AGS background;
additional muon loss detectors, which improved the study of time
dependence;
and further refined analyses, in particular of coherent betatron
oscillations.

\section{Data Analysis}
The analysis of $a_\mu$ follows, naturally, the separation of the
measurement in the frequencies $\omega_p$ and $\omega_a$.
Both frequencies were analyzed independently by several groups
within the collaboration.
The magnetic field frequencies measured during the running period were
weighted by the distribution of analyzed muons, both in time and over
the storage region.
The frequency fitted from the positron time spectra was corrected by
+0.76(3)\,ppm for the net contribution to the muon spin precession and
momentum rotation caused by vertical beam oscillations and,
for muons with $\gamma \neq 29.3$, by horizontal electric fields~\cite{farley}.
The values of $R = \omega_a/\omega_p$ and $a_\mu$ were evaluated only
after each of the frequency analyses had been finalized; at no earlier
stage were the absolute values of both frequencies, $\omega_p$ and
$\omega_a$, known to any of the collaborators.

\subsection{The frequency $\omega_p$}
The analysis of the magnetic field data starts with the calibration of
the 17 NMR probes in the field trolley using dedicated measurements taken
during and at the end of the data collection period.
In these calibration measurements, the field in the storage region was
tuned to very good homogeneity at two specific calibration locations.
The field was then measured with the NMR probes mounted in the
trolley shell, as well as with a single probe plunged into the storage
vacuum and positioned to measure the field values in the corresponding
locations.
Drifts of the field during the calibration measurements were determined
by remeasuring the field with the trolley after the measurements with the
plunging probe were completed, and in addition by interpolation of the
readings from nearby NMR probes in the outer top and bottom walls of the
vacuum chamber.
The difference of the trolley and plunging probe readings forms
a calibration of the trolley probes with respect to the plunging probe,
and hence with respect to each other.
The plunging probe, as well as a subset of the trolley probes, were
calibrated with respect to a standard probe~\cite{fei} at the end of
the running period in a similar sequence of measurements in the storage
region, which was opened to air for that purpose.
The leading uncertainties in the calibration procedure result from the
residual inhomogeneity of the field at the calibration locations, and
from position uncertainties in the active volumes of the NMR probes.
These uncertainties were evaluated from measurements in which the
trolley shell was purposely displaced and known field gradients were
applied using the so-called surface and dipole correction coils of the
ring magnet.
The size of these uncertainties is estimated to be 0.15\,ppm, as listed
in Table~\ref{table:field}.
The uncertainty in the absolute calibration of the standard probe amounts
to 0.05\,ppm~\cite{fei}.
The dependencies of the trolley NMR readings on the supply voltage and on
other parameters were measured to be small in the range of operation.
An uncertainty of 0.10\,ppm ("Others" in Table~\ref{table:field}) is
assigned, which includes also the measured effects from the transient
kicker field caused by eddy currents and from AGS stray fields.
\begin{table}
\caption {Systematic uncertainties for the $\omega_p$ analysis.  The uncertainty "Others" groups uncertainties caused by higher multipoles, the trolley frequency, temperature, and voltage response, eddy currents from the kickers, and time-varying stray fields.}
\begin{center}
\begin{tabular}{l|c}
\hline
Source of errors & Size [ppm] \\
\hline
Absolute calibration of standard probe\hspace{3em} & 0.05\\
Calibration of trolley probe & 0.15\\
Trolley measurements of $B_0$ & 0.10\\
Interpolation with fixed probes & 0.10\\
Uncertainty from muon distribution & 0.03\\
Others & 0.10\\
\hline
Total systematic error on $\omega_p$ & 0.24 \\
\hline
\end{tabular}
\end{center}
\label{table:field}
\end{table}

\begin{figure}
  \begin{center}
    \includegraphics[width=\textwidth]{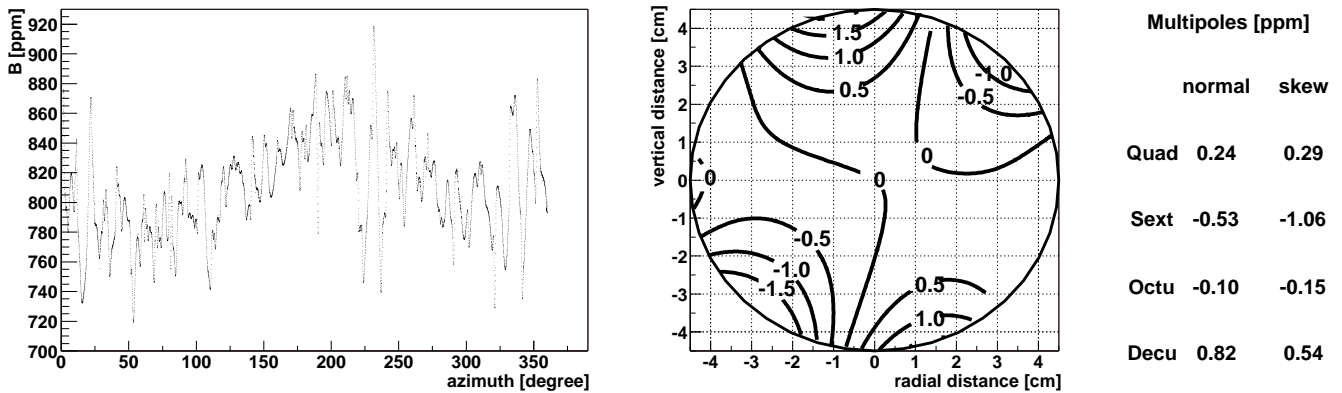}
  \end{center}
  \caption{The NMR frequency measured with the center trolley probe relative
  to a 61.74\,MHz reference versus the azimuthal position in the storage ring
  (left), and (right) a 2-dimensional multipole expansion of theazimuthal
  average of the field measured with 15 trolley probes with respect to the
  central field value of 1.451\,275\,T.
  The multipole amplitudes are given at the storage ring aperture, which has
  a 4.5\,cm radius as indicated by the circle.}
  \label{fig:field}
\end{figure}

The magnetic field inside the storage region was measured 22 times with
the field trolley during the data collection from January to March 2000.
Fig.~\ref{fig:field}a shows the field value measured in the storage ring with
the center trolley probe versus the azimuthal angle.
The field is seen to be uniform to within about $\pm 50\,$ppm of its average
value over the full azimuthal range, including the  region near $350^\circ$
where the inflector magnet is located.
Non-linearities in the determination of the trolley position during the
measurements --- from the measured cable lengths and from perturbations
on the readings from fixed probes as the trolley passes --- are estimated
to affect the azimuthal average of the field at the level of 0.10\,ppm.
Fig.~\ref{fig:field}b shows a 2-dimensional multipole expansion of the
azimuthally averaged readings from the trolley probes,
\begin{eqnarray}
  B_y & = & \sum_{n=0}^{\infty} C_n r^n \cos(n\phi) - 
            \sum_{n=0}^{\infty} D_n r^n \sin(n\phi), \\
  B_x & = & \sum_{n=0}^{\infty} C_n r^n \sin(n\phi) +
            \sum_{n=0}^{\infty} D_n r^n \cos(n\phi)
,
\end{eqnarray}
where the coefficients $C_n$ and $D_n$ are the normal and skew multipoles,
and $r$ and $\phi$ denote the polar coordinates in the storage region.
The multipole expansion was truncated in the analysis
after the decupoles.
Measurements with probes extending to larger radii
show that the neglect of higher multipoles is at most 0.03\,ppm in terms
of the  average field encountered by  the stored muons,
in agreement with magnet design calculations.
The field averaged over azimuth is seen to be uniform to within 1.5\,ppm of
its value.

The measurements with the trolley relate
the readings of 370 NMR fixed probes in the outer top and bottom walls
of the storage vacuum chamber to the field values in the beam region.
The fixed NMR probes are used to interpolate the field when the field trolley
is 'parked' in the storage vacuum just outside the beam region, and muons
circulate in the storage ring.
Since the relationship between the field value in the storage region and
the fixed probe readings may change during the course of the data collection
period,
the field mappings with the trolley were repeated typically two to three
times per week, and whenever ramping of the magnet or a change in
settings required such.
The uncertainty associated with the interpolation of the
magnetic field between trolley measurements is estimated from the spread
of the difference between the dipole moments evaluated from the fixed probe
measurements and from the trolley probe measurements in periods of constant
magnet settings and powering.  It is found to be 0.10\,ppm.

Since the field is highly uniform, the field integral encountered by the
(analyzed) muons is rather insensitive to the exact location of the beam.
As in earlier works~\cite{Brown:2001mg,Brown:2000sj}, the radial
equilibrium beam position was determined from the debunching of the beam
following injection and the vertical position from the distribution of counts
in scintillation counters mounted on the front faces of the positron
calorimeters.
The position uncertainty amounts to 1 -- 2\,mm, which contributes
0.03\,ppm uncertainty to the field integral.

The result for field frequency $\omega_p$ weighted by the muon distribution
is found to be,
\begin{equation}
  \omega_p/(2\pi) = 61\,791\,595(15)\,\mathrm{Hz}~~\mbox{(0.2\,ppm)},
  \label{eq:omegap_result}
\end{equation}
where the uncertainty has a leading contribution from the calibration of
the trolley probes and is thus predominantly systematic.
A second analysis of the field has been performed using additional
calibration data, a different selection of fixed NMR probes, and
a different method to relate the trolley and fixed probe readings.
The results from these analyses are found to agree to within a fraction
of the total uncertainty on $\omega_p$.

\subsection{The frequency $\omega_a$}
\begin{figure}
    \includegraphics[width=\textwidth]{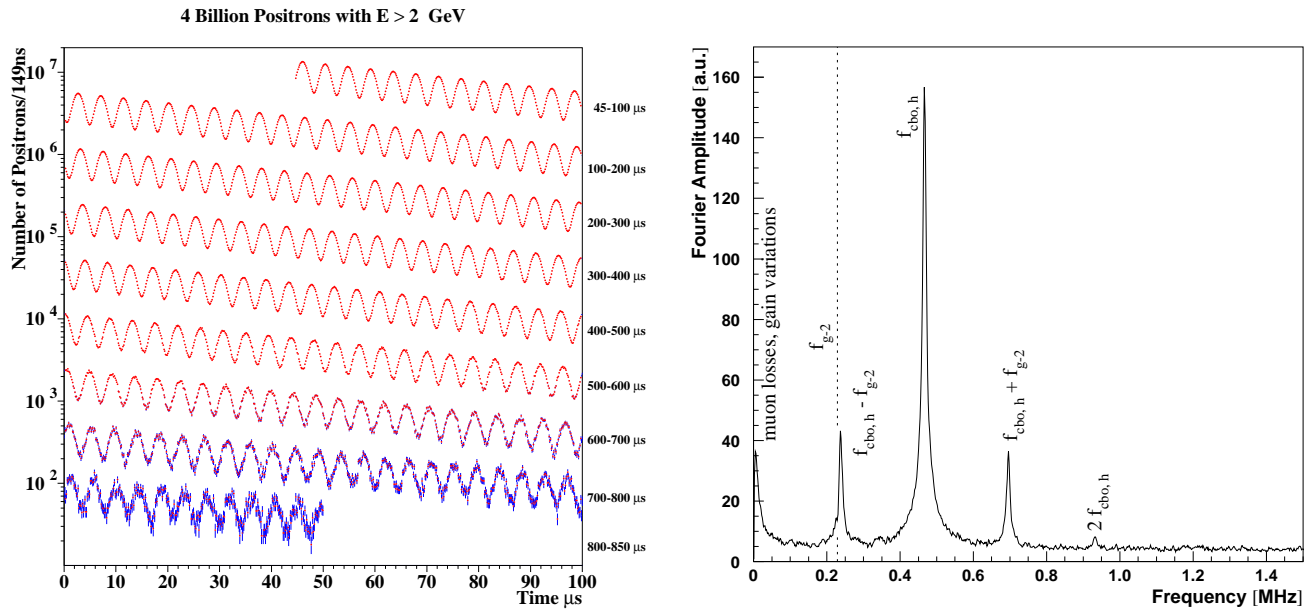}
  \caption{The time spectrum for $4 \cdot 10^9$ positrons with energies greater than 2\,GeV collected from January to March 2000, after corrections for pile-up and for the bunched time structure of the injected beam (left) were made, and (right) the Fourier transform of the time spectrum, in which muon decay and spin precession (cf. Eq.~2) has been suppressed to emphasize other effects.
  }
  \label{fig:precession}
\end{figure}

The event sample available for analysis from data collection in the year
2000 amounts to about $4\cdot10^9$ positrons  reconstructed with energies
greater than 2\,GeV  and times  between  \mbox{50\,$\mu$s} and
\mbox{600\,$\mu$s} following the injection of a beam bunch.
Fig.~\ref{fig:precession}a shows the time spectrum corrected for
the bunched time structure of the beam and for overlapping calorimeter
pulses, so called pile-up~\cite{Brown:2001mg}.

The leading characteristics of the time spectrum are those of muon decay
and spin precession (cf. Eq.~\ref{eq:precession}).
Additional effects exist, as seen from the Fourier spectrum in
Fig.~\ref{fig:precession}b, and require careful consideration in the
analysis.
These effects include detector gain and time instability, muon losses,
and oscillations of the beam as a whole, so-called coherent betatron
oscillations (CBO).

Numerically most relevant to the determination of $\omega_a$ are
CBO in the horizontal plane.
CBO are caused by injecting the beam through the relatively narrow
$18(\rm{w}) \times 57(\rm{h})\,\rm{mm}^2$ aperture of the 1.7\,m long
inflector channel into the 90\,mm diameter aperture of the storage region,
and have been observed directly with fiber harp monitors plunged into the beam
region for this purpose.
The CBO frequency is determined by the focusing index of the storage ring,
and is numerically close to twice the frequency $\omega_a$ for the
quadrupole settings employed in most measurements so far.
Since the calorimeter acceptances vary with the radial muon decay position
in the storage ring and with the momentum of the decay positron, the
time and energy spectra of the observed positrions are modulated with
the CBO frequency.
These modulations affect the normalization $N_0$, the asymmetry $A$, and
the phase $\phi$ in Eq.~\ref{eq:precession} at the level of 1\%, 0.1\%,
and 1\,mrad at beam injection.
When not accounted for in the function fitted to the data, the
modulations of the asymmetry and phase with a frequency
$\omega_\mathrm{cbo,h} \simeq 2 \times \omega_a$
may manifest themselves as artificial shifts of up to 4\,ppm in the
frequency values $\omega_a$ determined from individual calorimeter
spectra.
The circular symmetry of the experiment design results in a strong
cancellation of such shifts in the joined calorimeter spectrum.

Several approaches have been pursued in the analysis of $\omega_a$.
In one approach, the time spectra from individual positron calorimeters
was fitted in narrow energy intervals using a fit function as in
Eq.~\ref{eq:precession} extended by the number, asymmetry, and phase
modulations.
Other approaches made use of the cancellation in the joined calorimeter
spectra and either fitted for the residual of the leading effects, or
accounted for their neglect in a contribution to the systematic uncertainty.
The results are found to agree, on $\omega_a$ to within the expected 0.5\,ppm
statistical variation resulting from the slightly different
selection and treatment of the data in the respective analyses.
The combined result is found to be,
\begin{equation}
  \omega_a/(2\pi) = 229\,074\,11(14)(7)\,\mathrm{Hz}~~\mbox{(0.7\,ppm)},
\end{equation}
in which the first uncertainty is statistical and the second systematic.
The combined systematic uncertainty is broken down by source in
Table~\ref{table:precession}.
\begin{table}
\caption{Systematic uncertainties for the $\omega_a$ analysis.  The uncertainty "Others" groups uncertainties caused by AGS background, timing shifts, vertical oscillations and radial electric fields, and beam debunching/randomization.}
\begin{center}
\begin{tabular}{l|c}
\hline
Source of errors & Size [ppm] \\
\hline
Coherent betatron oscillations\hspace{6em}  &  0.21\\
Pileup & 0.13\\
Gain changes & 0.13\\
Lost muons & 0.10\\
Binning and fitting procedure & 0.06\\
Others & 0.06\\
\hline
Total systematic error on $\omega_a$ & 0.31\\
\hline
\end{tabular}
\end{center}
\label{table:precession}
\end{table}

\section{Results and Discussion}
\begin{figure}
  \begin{center}
    \includegraphics[width=0.60\textwidth]{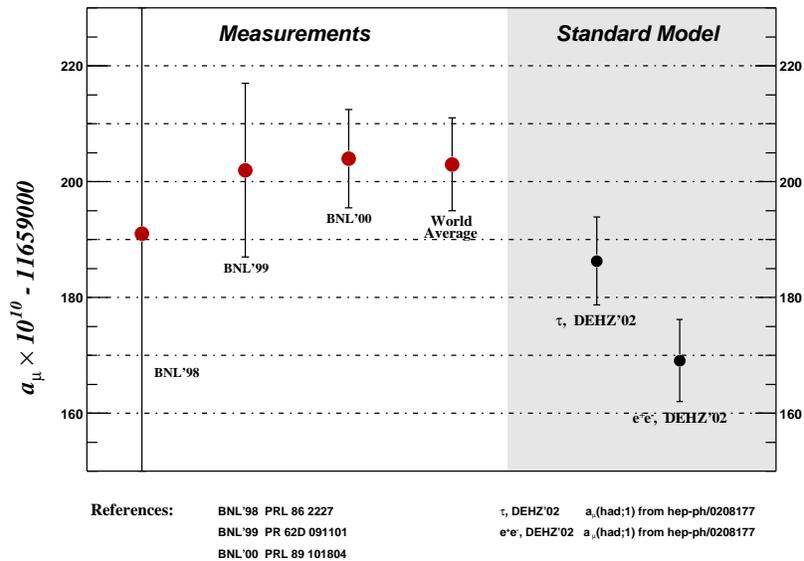}
  \end{center}
  \caption{Recent measuremens of $a_\mu$ and standard model evaluations
  using the evaluations in Ref.~\cite{Davier:2002dy} of the lowest order
  contribution from hadronic vacuum polarization.}
  \label{fig:result}
\end{figure}

The value of $a_\mu$ was evaluated after the analyses of
$\omega_p$ and $\omega_a$ had been finalized,
\begin{equation}
  a_{\mu^+} = 11\,659\,204(7)(5)\ \times\ 10^{-10}~~\mbox{(0.7\,ppm)},
\end{equation}
where the first uncertainty is statistical and the second systematic.
This new result is in good agreement with the previous
measurements~\cite{cern-3,Brown:2001mg,Brown:2000sj,Carey:1999dd} and
drives the present world average,
\begin{equation}
  a_\mu(\mathrm{exp}) = 11\,659\,203(8)\ \times\ 10^{-10}~~\mbox{(0.7\,ppm)},
\end{equation}
in which the uncertainty accounts for known correlations between the
systematic uncertainties in the measurements.
Fig.~\ref{fig:result} shows our recent measurements of $a_{\mu^+}$,
together with two SM evaluations discussed below.

In the SM, the value of $a_\mu$ receives contributions
from QED, hadronic, and electroweak processes,
$a_\mu(\mathrm{SM}) =
  a_\mu(\mathrm{QED})  + a_\mu(\mathrm{had}) + a_\mu(\mathrm{weak})$.
The QED and weak contributions can, unlike the hadronic contribution, be
evaluated perturbatively,
$a_\mu(\mathrm{QED}) = 11\,658\,470.57(29)\,\times\,10^{-10}$~\cite{Mohr:2000ie}
and $a_\mu(\mathrm{weak}) = 15.1(4)\,\times\,10^{-10}$~\cite{Czarnecki:2001pv}.
The hadronic contribution is, in lowest order, related by dispersion theory
to the hadron production cross sections measured in $e^+e^-$ collisions
and, under additional assumptions, to hadronic $\tau$-decay.
Clearly, the hadronic contribution has a long history of values
as new data appeared and analyses were refined.

Shortly before the SPIN-2002 conference, Davier and co-workers released a
new and detailed evaluation~\cite{Davier:2002dy},
which now incorporates the high precision $e^+e^-$
data~\cite{Akhmetshin:2001ig} in the region of the $\rho$ resonance
from CMD-2 at Novosibirsk,
more accurate $e^+e^-$ measurements~\cite{Bai:1999pk,Bai:2001ct}
in the 2--5\,GeV energy region from BES in Beijing,
preliminary results from the final ALEPH analysis~\cite{aleph} of hadronic
$\tau$-decay at LEP1, as well as additional CLEO
data~\cite{Anderson:1999ui,Edwards:1999fj}.
The authors note discrepancies between the $e^+e^-$ and $\tau$ data at
the present levels of precision, and obtain separate predictions for the
contribution to $a_\mu(\mathrm{SM})$ from lowest order hadronic vacuum
polarization,
$a_\mu(\mathrm{had,1}) = 685(7)\,\times\,10^{-10}$ from $e^+e^-$ data and
$a_\mu(\mathrm{had,1}) = 702(6)\,\times\,10^{-10}$ from $\tau$ data.
Higher order contributions include higher order hadronic vacuum
polarization~\cite{Krause:1997rf,Alemany:1998tn} and
hadronic light-by-light
scattering~\cite{Hayakawa:2001bb,Bijnens:2001cq,Knecht:2001qf,Knecht:2001qg,Blokland:2001pb}.

Open questions concern the SM value of $a_\mu$, in particular the hadronic contribution,
and the experimental value of $a_{\mu^-}$ at sub-ppm precision.
The former should benefit from further theoretical scrutiny, e.g. Refs.~\cite{Kinoshita:2002ns, Czarnecki:2002nt, Hagiwara:2002ma, Erler:2002mv}, from
radiative-return measurements at $e^+e^-$ factories, e.g. Refs.~\cite{Venanzoni:2002jb,Solodov:2002xu}, and possibly from latice calculation~\cite{Blum:2002ii}.
We are currently analyzing a sample of about \mbox{$3\,\times\,10^9$}
decay electrons from $a_{\mu^-}$ data collected in the year 2001.
Stay tuned!

\section*{Acknowledgements}
This work was supported in part by the U.S. Department of Energy, the U.S. National Science Foundation, the U.S. National Computational Science Alliance, the German Bundesminister f\"{u}r Bildung und Forschung, the Russian Ministry of Science, and the U.S.-Japan Agreement in High Energy Physics.

\bibliography{sichtermann}

\end{document}